\documentclass[aps,prb,twocolumn]{revtex4}
\usepackage{graphicx}
\usepackage{dcolumn}
\usepackage{bm}
\usepackage{subfigure}
\usepackage{color}
\usepackage{hyperref}

\begin{document}

\title{Analysis of Possible Quantum Metastable States in Ballistic Graphene-based Josephson Junctions}

\author{Joseph G. Lambert}
\email{jgl29@drexel.edu}
\author{Steve Carabello}
\author{Roberto C. Ramos}
\affiliation{Department of Physics, Drexel University, Philadelphia, PA 19104 USA}

\homepage{http://www.physics.drexel.edu/research/lowtemp/}

\date{\today}

\begin{abstract}
Graphene is a relatively new material (2004) made of atomic layers of carbon arranged in a honeycomb lattice. Josephson junction devices are made from graphene by depositing two parallel superconducting leads on a graphene flake. These devices have hysteretic current-voltage characteristics with a supercurrent branch and Shapiro steps appear when irradiated with microwaves. These properties motivate us to investigate the presence of quantum metastable states similar to those found in conventional current-biased Josephson junctions. We present work investigating the nature of these metastable states for ballistic graphene Josephson junctions. We model the effective Washboard potential for these devices and estimate parameters, such as energy level spacing and critical currents, to deduce the design needed to observe metastable states. We propose devices consisting of a parallel on-chip capacitor and suspended graphene. The capacitor is needed to lower the energy level spacing down to the experimentally accessible range of 1-20 GHz. The suspended graphene helps reduce the noise that may otherwise come from two-level states in the insulating oxide layer. Moreover, back-gate voltage control of its critical current introduces another knob for quantum control. We will also report on current experimental progress in the area of fabrication of this proposed device.
\end{abstract}


\maketitle
\section{Introduction}

Since its isolation in 2004 \cite{Novoselov2004}, the electronic properties of graphene have inspired a flourishing field of research. Graphene is a two-dimensional, hexagonal lattice of carbon atoms that effectively contains relativistic Dirac charge carriers. Its characteristic linear dispersion relation where the upper and lower energy bands meet \cite{Novoselov2005} has led to novel and interesting physics and applications.

Recent studies \cite{Heersche2007, Heersche200772, HeerscheEPJ2007, Du2008, Miao2009} have demonstrated the robust and reproducible phenomenon of the Josephson effect in devices consisting of two superconducting leads contacted by flakes of graphene that are 1-4 layers thick. Clear evidence of multiple Andreev reflections \cite{HeerscheEPJ2007, Du2008} illustrates that such graphene junctions behave as SNS junctions. The critical current can be tuned by modulating a back-gate voltage \cite{Heersche2007, Miao2009}. A dc-SQUID has been constructed using graphene \cite{Girit2009}, demonstrating several advantages.  These include how graphene is straightforward to fabricate, and electrically contact with high-transparency electrodes. The current-phase relation has been predicted \cite{Girit2009CPR} and measured \cite{Chialvo2010}.

The appearance of the Josephson effect in graphene junstions motivates us to investigate potential metastable quantum states within the washboard potential wells.  Such states have been extensively studied in conventional junctions for over 25 years and have been used as basis states for superconducting quantum computing \cite{Martinis1985}.

\section{Device Fabrication and Characterization}

\begin{figure}[htbp]
		\includegraphics[scale=.28]{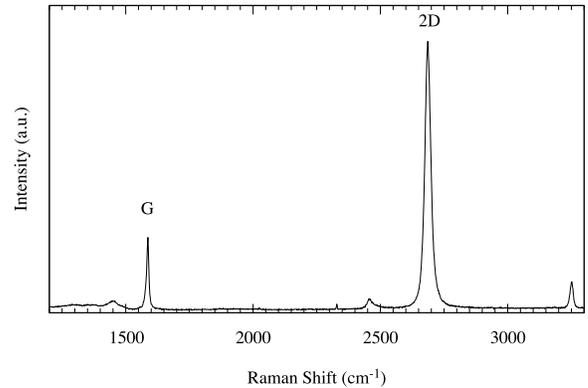}
\caption{\label{raman}Example Raman spectrum of our single-layer graphene specimens using 514.5 nm laser excitation. The G peak at $\sim$ 1580 cm$^{-1}$ is the E$_{2g}$ optical mode due to in-plane lattice vibrations and the 2D peak at $\sim$ 2700 cm$^{-1}$ is the over tone of the D peak due to second-order double resonance \cite{Ferrari2006}. The ratio of the peak heights, and the positions of the peaks are signatures of the thickness of the graphene flake. }

\end{figure}

To fabricate devices, we prepare graphene using the mechanical exfoliation technique onto Si/SiO$_2$ (300nm) substrates \cite{Novoselov2004}.  We first locate and identify single-layer and multi-layer graphene specimens using optical microscopy \cite{blake2007}.  We confirm and characterize the number of layers using Raman spectroscopy \cite{Ferrari2006}.  An example of a typical Raman spectrum that we collect for single-layer graphene is shown in Fig. \ref{raman}.

\begin{figure}[htbp]
  	\begin{center}
		\subfigure[]{\label{fig2a}\includegraphics[width=3.3in]{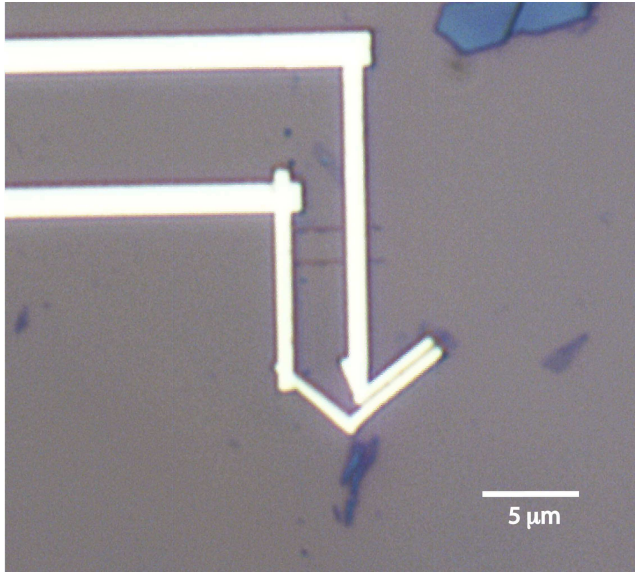}}
		\subfigure[]{\label{fig2b}\includegraphics[width=3.3in]{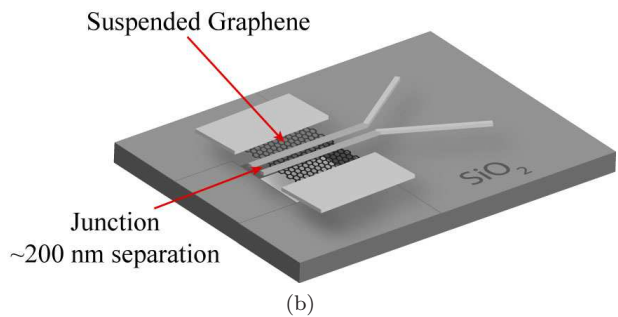}}
	\end{center}
\caption{ \label{fig:device}(a) Optical image of one of our fabricated graphene devices that is not suspended.  The lead separation is $L = 200$ nm and the length of the leads in contact with graphene is $W = 700$ nm.  (b) Schematic of our future graphene device with suspended graphene.  The graphene and leads are elevated from the substrate only for visual clarity.  }

\end{figure}

For device fabrication, we perform no Raman spectroscopy to prevent laser-induced defects in the graphene.  We fabricate junctions by depositing, onto the graphene flake, Ti (7nm)/Al (80nm) parallel leads, which are separated by 100 nm to 500 nm.  The Ti layer provides transparent adhesion to the graphene, and the Al layer is superconducting below $\sim$1 K.  Due to the superconducting proximity effect \cite{Heersche2007, Heersche200772}, the areas of graphene underneath the leads become superconducting.  We have fabricated devices, an example of which is depicted in Fig. \ref{fig2a}.  This device has a room temperature, 2-probe resistance of 5.8 k$\Omega$, which is evidence of good electrical contact.

Charge carrier mobility in graphene is reduced by substrate induced scatterers, such as charges trapped under the graphene and rippling of the graphene due to the roughness of the substrate \cite{Chen2009}. Removal of the SiO$_2$ substrate should also reduce the effects of anomalous two-level systems that plague superconducting qubits \cite{Cooper2004, Martinis2005, Palomaki2010}. In order to achieve ballistic transport in graphene, future devices will consist of suspended graphene, as in Fig. \ref{fig2b} \cite{Bolotin2008,Du2009,Bolotin2009}.  Suspension of the graphene will be achieved by etching the SiO$_2$ with hydrofluoric acid (HF).  HF solution is targeted underneath the graphene by capillary action \cite{Du2008susg}.  

The back-gate is provided by the doped Si substrate. For the case of non-suspended graphene, the back-gate and graphene are separated by the 300 nm of SiO$_2$ as an insulating dielectric, and for suspended graphene, they are separated by vacuum.  The on-chip shunting capacitor will have an interdigitated design, which will be fabricated with the device leads during a single deposition.

\section{Theory}

The quantum metastable states we are investigating are analogous to those existing within the wells of the tilted washboard potential of a current-biased Josephson junction \cite{919517, Xuthesis}.  This potential has the form

\begin{eqnarray}
\label{convwashboard}
U(\gamma)=-\frac{\Phi_0}{2\pi}\left(I_0\cos\left(\gamma\right)+I_b\gamma\right)
\end{eqnarray}

\noindent where $\gamma$ is the gauge-invariant phase difference, $\Phi_0=h/2\pi$ is the flux quantum, $I_0$ is the critical current, and $I_b$ is the bias current.  For a junction with capacitance $C$, this system is analogous to a Òphase particleÓ of mass $m = C(\Phi_0/2\pi)^2$ oscillating in a local well with plasma frequency \cite{Xuthesis}

\begin{eqnarray}
\label{convplasma}
\omega_p=\omega_0\left[1-\left(\frac{I_b}{I_0}\right)^2\right]^\frac14
\end{eqnarray}

\noindent where $\omega_0=\sqrt{2\pi I_0/\Phi_0 C}$.  At bias current $I_b\sim I_0$, the energy spacing between the ground and the first excited states is $\hbar\omega_{01}\approx\hbar\omega_p$ \cite{Xuthesis}.

The current-phase relationship, the critical current, and $I_0 R_N$ product for ballistic graphene junctions were calculated by Titov and Beenakker \cite{Titov2006}. Specifically, at the Dirac point, these equations are analytic:

\begin{eqnarray}
\label{eqn:ballisticCPR}
I(\gamma)&=&\frac{e\Delta_0}{\hbar}\frac{2W}{\pi L}\cos(\gamma/2)\rm{arctanh}\left[\sin(\gamma/2)\right]\\
\label{eqn:ballisticCC}
I_0&=&1.33\frac{e\Delta_0}{\hbar}\frac{W}{\pi L}\\
\label{eqn:ballisticI0Rn}
I_0 R_N&=&2.08\frac{\Delta_0}{e}.
\end{eqnarray}

\noindent Here, $\Delta_0$ is the superconducting energy gap of the leads, $W$ is the length of the leads in contact with the graphene flake, and $L$ is the lead separation.  These equations were derived assuming $L \ll W, \xi$ where $\xi = \hbar v/\Delta_0$ is the superconducting coherence length, and $v$ is the velocity of charge carriers.
According to the RCSJ model, the equation of motion of the ``phase particle'' is \cite{Xuthesis}

\begin{eqnarray}
\label{eqn:motion}
I_b=I(\gamma)+C\frac{\Phi_0}{w\pi}\ddot{\gamma}+\frac{\Phi_0}{2\pi R}\dot{\gamma}.
\end{eqnarray} 

Substituting (\ref{eqn:ballisticCPR}) in for $I(\gamma)$ and working backwards, we obtain a washboard potential for a ballistic superconductor-graphene-superconductor (SGS) junction:

\begin{widetext}
\begin{eqnarray}
\label{eqn:ballisticWBpot}
U(\gamma)=-\frac{\Phi_0}{2\pi} \left(  -\frac{2I_0}{1.33}\left[2\sin\left(\frac\gamma2\right)\rm{arctanh}\left(\sin\left(\frac\gamma2\right)\right)+\ln\left(1-\sin^2\left(\frac\gamma2\right)\right)\right]+I_b\gamma  \right).
\end{eqnarray}
\end{widetext}

\begin{figure}[htbp]
  	\begin{center}
		\subfigure[]{\label{fig3a}\includegraphics[width=3in]{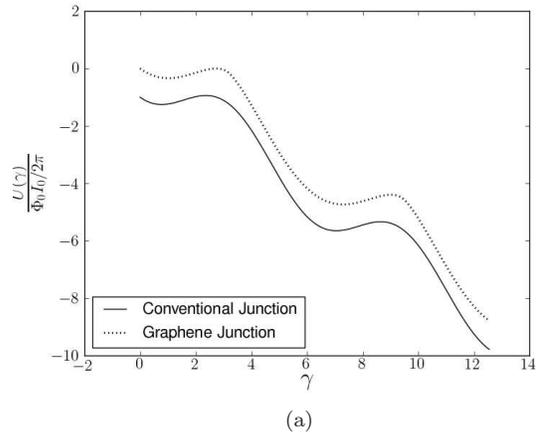}}
		\subfigure[]{\label{fig3b}\includegraphics[width=3in]{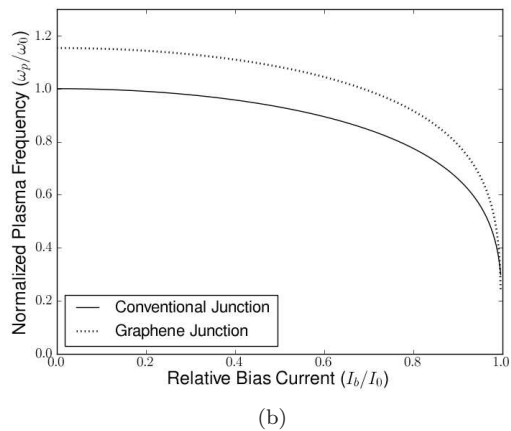}}
	\end{center}
\caption{ \label{fig:plots}(a) Plot of the ballistic graphene junction washboard potential biased at $i_b=I_b/I_0=0.5$ compared to the conventional washboard potential.  The vertical axis is in units of $\Phi_0 I_0/2\pi$.  (b)  Comparison of $f$ vs. $I$ for conventional and ballistic graphene junctions.  }

\end{figure}

The SGS potential (\ref{eqn:ballisticWBpot}) is plotted alongside the conventional washboard potential (\ref{convwashboard}) in Fig. \ref{fig3a}. For the same critical current and capacitance, the wells of the SGS potential are shallower.  This would result in more closely spaced energy levels.

From (\ref{eqn:ballisticWBpot}) we calculate the plasma frequency about the local minimum of the well, $\gamma_{\text{min}}$, for ballistic graphene junctions.  This is plotted and compared to (2) in Fig. 3(b).

\begin{eqnarray}
\label{eqn:ballisticplasma}
\omega_p&=& \left(\frac{2\pi I_0}{1.33\Phi_0 C}\right)^{\frac{1}2}\nonumber\\
&\times&\left[1-\sin\left(\frac{\gamma_{\text{min}}}{2}\right)\tanh^{-1}\left(\sin\left(\frac{\gamma_{\text{min}}}{2}\right)\right)\right]^{\frac12}.
\end{eqnarray}

\vspace{.1in}

	The superconducting gap for bulk Al is $\Delta_0=180$ $\mu$eV; however, experimentally determined values \cite{Miao2009} in typical devices ranged from $90$ to $120$ $\mu$eV.  Using the approximation $\Delta_0\approx100$ $\mu$eV, we estimate a critical current of $I_0 \approx 100$ nA for the device shown in Fig. \ref{fig2a} $(L=200\text{ nm,} W=700\text{ nm})$.  The intrinsic capacitance of the device is small, $\sim10^{-16}$  F.  Using the estimated critical current and (\ref{convplasma}) gives $f_p=\omega_p/2\pi \approx 300$ GHz, in qualitative agreement with experimental results \cite{Du2008, Girit2009}.  We can lower this frequency to $f_p \approx 5$ GHz by adding a $0.1$ pF shunting capacitor.

\section{Conculsion}
We have calculated the washboard potential and plasma frequency of SGS junctions.  Our calculations show that ballistic graphene junctions are essentially similarly to conventional junctions, with the added benefit of a back-gate voltage control of the critical current.  We are testing this conclusion with a series of experiments such as current-voltage measurements, microwave resonant activation, and quantum tunneling \cite{washburn1985} experiments.  We will also measure the graphene junctions in both the classical and quantum regimes. The crossover temperature, $T_c=\hbar\omega/2\pi k_B$, distinguishes these two regimes \cite{Tinkham}.  At current biasing near the critical current, the estimated plasma frequency is $100$ GHz, and with a shunting capacitor of $0.1$ pF, it is $5$ GHz.   These plasma frequencies have crossover temperatures of approximately $800$ mK and $40$ mK, respectively.   For the higher plasma frequencies, multi-photon processes will be used to enhance the escape of the phase particle \cite{Wallraff2003}.  We will tune these experiments with the back-gate voltage to study this unique graphene/superconductor system.

\begin{acknowledgments}
We acknowledge Prof. Yury Gogotsi and Min Heon (Drexel Nanomaterials Group) for providing high-quality graphite and Prof. Jonathan Spanier and Eric Gallo (Drexel Mesomaterials Group) for help with fabrication. We also thank Prof. Michael Fuhrer's group at the University of Maryland for helpful discussions and advice about graphene.  Devices were fabricated at the Wolf Nanofabrication Facilities at the University of Pennsylvania.
\end{acknowledgments}

\bibliography{ref}

\begin{thebibliography}{27}
\expandafter\ifx\csname natexlab\endcsname\relax\def\natexlab#1{#1}\fi
\expandafter\ifx\csname bibnamefont\endcsname\relax
  \def\bibnamefont#1{#1}\fi
\expandafter\ifx\csname bibfnamefont\endcsname\relax
  \def\bibfnamefont#1{#1}\fi
\expandafter\ifx\csname citenamefont\endcsname\relax
  \def\citenamefont#1{#1}\fi
\expandafter\ifx\csname url\endcsname\relax
  \def\url#1{\texttt{#1}}\fi
\expandafter\ifx\csname urlprefix\endcsname\relax\def\urlprefix{URL }\fi
\providecommand{\bibinfo}[2]{#2}
\providecommand{\eprint}[2][]{\url{#2}}

\bibitem[{\citenamefont{Novoselov et~al.}(2004)\citenamefont{Novoselov, Geim,
  Morozov, Jiang, Zhang, Dubonos, Grigorieva, and Firsov}}]{Novoselov2004}
\bibinfo{author}{\bibfnamefont{K.~S.} \bibnamefont{Novoselov}},
  \bibinfo{author}{\bibfnamefont{A.~K.} \bibnamefont{Geim}},
  \bibinfo{author}{\bibfnamefont{S.~V.} \bibnamefont{Morozov}},
  \bibinfo{author}{\bibfnamefont{D.}~\bibnamefont{Jiang}},
  \bibinfo{author}{\bibfnamefont{Y.}~\bibnamefont{Zhang}},
  \bibinfo{author}{\bibfnamefont{S.~V.} \bibnamefont{Dubonos}},
  \bibinfo{author}{\bibfnamefont{I.~V.} \bibnamefont{Grigorieva}},
  \bibnamefont{and} \bibinfo{author}{\bibfnamefont{A.~A.}
  \bibnamefont{Firsov}}, \bibinfo{journal}{Science}
  \textbf{\bibinfo{volume}{306}}, \bibinfo{pages}{666} (\bibinfo{year}{2004}).

\bibitem[{\citenamefont{Novoselov et~al.}(2005)\citenamefont{Novoselov, Geim,
  Morozov, Jiang, Katsnelson, Grigorieva, Dubonos, and Firsov}}]{Novoselov2005}
\bibinfo{author}{\bibfnamefont{K.~S.} \bibnamefont{Novoselov}},
  \bibinfo{author}{\bibfnamefont{A.~K.} \bibnamefont{Geim}},
  \bibinfo{author}{\bibfnamefont{S.~V.} \bibnamefont{Morozov}},
  \bibinfo{author}{\bibfnamefont{D.}~\bibnamefont{Jiang}},
  \bibinfo{author}{\bibfnamefont{M.~I.} \bibnamefont{Katsnelson}},
  \bibinfo{author}{\bibfnamefont{I.~V.} \bibnamefont{Grigorieva}},
  \bibinfo{author}{\bibfnamefont{S.~V.} \bibnamefont{Dubonos}},
  \bibnamefont{and} \bibinfo{author}{\bibfnamefont{A.~A.}
  \bibnamefont{Firsov}}, \bibinfo{journal}{Nature}
  \textbf{\bibinfo{volume}{438}}, \bibinfo{pages}{197} (\bibinfo{year}{2005}).

\bibitem[{\citenamefont{Heersche
  et~al.}(2007{\natexlab{a}})\citenamefont{Heersche, Jarillo-Herrero, Oostinga,
  Vandersypen, and Morpurgo}}]{Heersche2007}
\bibinfo{author}{\bibfnamefont{H.~B.} \bibnamefont{Heersche}},
  \bibinfo{author}{\bibfnamefont{P.}~\bibnamefont{Jarillo-Herrero}},
  \bibinfo{author}{\bibfnamefont{J.~B.} \bibnamefont{Oostinga}},
  \bibinfo{author}{\bibfnamefont{L.~M.~K.} \bibnamefont{Vandersypen}},
  \bibnamefont{and} \bibinfo{author}{\bibfnamefont{A.~F.}
  \bibnamefont{Morpurgo}}, \bibinfo{journal}{Nature}
  \textbf{\bibinfo{volume}{446}}, \bibinfo{pages}{56}
  (\bibinfo{year}{2007}{\natexlab{a}}).

\bibitem[{\citenamefont{Heersche
  et~al.}(2007{\natexlab{b}})\citenamefont{Heersche, Jarillo-Herrero, Oostinga,
  Vandersypen, and Morpurgo}}]{Heersche200772}
\bibinfo{author}{\bibfnamefont{H.~B.} \bibnamefont{Heersche}},
  \bibinfo{author}{\bibfnamefont{P.}~\bibnamefont{Jarillo-Herrero}},
  \bibinfo{author}{\bibfnamefont{J.~B.} \bibnamefont{Oostinga}},
  \bibinfo{author}{\bibfnamefont{L.~M.} \bibnamefont{Vandersypen}},
  \bibnamefont{and} \bibinfo{author}{\bibfnamefont{A.~F.}
  \bibnamefont{Morpurgo}}, \bibinfo{journal}{Solid State Communications}
  \textbf{\bibinfo{volume}{143}}, \bibinfo{pages}{72 }
  (\bibinfo{year}{2007}{\natexlab{b}}).

\bibitem[{\citenamefont{Heersche
  et~al.}(2007{\natexlab{c}})\citenamefont{Heersche, Jarillo-Herrero, Oostinga,
  Vandersypen, and Morpurgo}}]{HeerscheEPJ2007}
\bibinfo{author}{\bibfnamefont{H.~B.} \bibnamefont{Heersche}},
  \bibinfo{author}{\bibfnamefont{P.}~\bibnamefont{Jarillo-Herrero}},
  \bibinfo{author}{\bibfnamefont{J.~B.} \bibnamefont{Oostinga}},
  \bibinfo{author}{\bibfnamefont{L.~M.~K.} \bibnamefont{Vandersypen}},
  \bibnamefont{and} \bibinfo{author}{\bibfnamefont{A.~F.}
  \bibnamefont{Morpurgo}}, \bibinfo{journal}{Eur. Phys. J. Special Topics}
  \textbf{\bibinfo{volume}{148}}, \bibinfo{pages}{27}
  (\bibinfo{year}{2007}{\natexlab{c}}).

\bibitem[{\citenamefont{Du et~al.}(2008{\natexlab{a}})\citenamefont{Du,
  Skachko, and Andrei}}]{Du2008}
\bibinfo{author}{\bibfnamefont{X.}~\bibnamefont{Du}},
  \bibinfo{author}{\bibfnamefont{I.}~\bibnamefont{Skachko}}, \bibnamefont{and}
  \bibinfo{author}{\bibfnamefont{E.~Y.} \bibnamefont{Andrei}},
  \bibinfo{journal}{Phys. Rev. B} \textbf{\bibinfo{volume}{77}},
  \bibinfo{pages}{184507} (\bibinfo{year}{2008}{\natexlab{a}}).

\bibitem[{\citenamefont{Miao et~al.}(2009)\citenamefont{Miao, Bao, Zhang, and
  Lau}}]{Miao2009}
\bibinfo{author}{\bibfnamefont{F.}~\bibnamefont{Miao}},
  \bibinfo{author}{\bibfnamefont{W.}~\bibnamefont{Bao}},
  \bibinfo{author}{\bibfnamefont{H.}~\bibnamefont{Zhang}}, \bibnamefont{and}
  \bibinfo{author}{\bibfnamefont{C.~N.} \bibnamefont{Lau}},
  \bibinfo{journal}{Solid State Communications} \textbf{\bibinfo{volume}{149}},
  \bibinfo{pages}{1046 } (\bibinfo{year}{2009}).

\bibitem[{\citenamefont{Girit et~al.}(2009{\natexlab{a}})\citenamefont{Girit,
  Bouchiat, Naaman, Zhang, Crommie, Zettl, and Siddiqi}}]{Girit2009}
\bibinfo{author}{\bibfnamefont{C.}~\bibnamefont{Girit}},
  \bibinfo{author}{\bibfnamefont{V.}~\bibnamefont{Bouchiat}},
  \bibinfo{author}{\bibfnamefont{O.}~\bibnamefont{Naaman}},
  \bibinfo{author}{\bibfnamefont{Y.}~\bibnamefont{Zhang}},
  \bibinfo{author}{\bibfnamefont{M.~F.} \bibnamefont{Crommie}},
  \bibinfo{author}{\bibfnamefont{A.}~\bibnamefont{Zettl}}, \bibnamefont{and}
  \bibinfo{author}{\bibfnamefont{I.}~\bibnamefont{Siddiqi}},
  \bibinfo{journal}{Nano Letters} \textbf{\bibinfo{volume}{9}},
  \bibinfo{pages}{198} (\bibinfo{year}{2009}{\natexlab{a}}).

\bibitem[{\citenamefont{Girit et~al.}(2009{\natexlab{b}})\citenamefont{Girit,
  Bouchiat, Naaman, Zhang, Crommie, Zettl, and Siddiqi}}]{Girit2009CPR}
\bibinfo{author}{\bibfnamefont{{\c{C}}.}~\bibnamefont{Girit}},
  \bibinfo{author}{\bibfnamefont{V.}~\bibnamefont{Bouchiat}},
  \bibinfo{author}{\bibfnamefont{O.}~\bibnamefont{Naaman}},
  \bibinfo{author}{\bibfnamefont{Y.}~\bibnamefont{Zhang}},
  \bibinfo{author}{\bibfnamefont{M.~F.} \bibnamefont{Crommie}},
  \bibinfo{author}{\bibfnamefont{A.}~\bibnamefont{Zettl}}, \bibnamefont{and}
  \bibinfo{author}{\bibfnamefont{I.}~\bibnamefont{Siddiqi}},
  \bibinfo{journal}{physica status solidi (b)} \textbf{\bibinfo{volume}{246}},
  \bibinfo{pages}{2568} (\bibinfo{year}{2009}{\natexlab{b}}).

\bibitem[{\citenamefont{Chialvo et~al.}(2010)\citenamefont{Chialvo, Moraru,
  Harlingen, and Mason}}]{Chialvo2010}
\bibinfo{author}{\bibfnamefont{C.}~\bibnamefont{Chialvo}},
  \bibinfo{author}{\bibfnamefont{I.~C.} \bibnamefont{Moraru}},
  \bibinfo{author}{\bibfnamefont{D.~J.~V.} \bibnamefont{Harlingen}},
  \bibnamefont{and} \bibinfo{author}{\bibfnamefont{N.}~\bibnamefont{Mason}}
  (\bibinfo{year}{2010}), \eprint{arXiv/1005.2630}.

\bibitem[{\citenamefont{Martinis et~al.}(1985)\citenamefont{Martinis, Devoret,
  and Clarke}}]{Martinis1985}
\bibinfo{author}{\bibfnamefont{J.~M.} \bibnamefont{Martinis}},
  \bibinfo{author}{\bibfnamefont{M.~H.} \bibnamefont{Devoret}},
  \bibnamefont{and} \bibinfo{author}{\bibfnamefont{J.}~\bibnamefont{Clarke}},
  \bibinfo{journal}{Phys. Rev. Lett.} \textbf{\bibinfo{volume}{55}},
  \bibinfo{pages}{1543} (\bibinfo{year}{1985}).

\bibitem[{\citenamefont{Blake et~al.}(2007)\citenamefont{Blake, Hill, Neto,
  Novoselov, Jiang, Yang, Booth, and Geim}}]{blake2007}
\bibinfo{author}{\bibfnamefont{P.}~\bibnamefont{Blake}},
  \bibinfo{author}{\bibfnamefont{E.~W.} \bibnamefont{Hill}},
  \bibinfo{author}{\bibfnamefont{A.~H.~C.} \bibnamefont{Neto}},
  \bibinfo{author}{\bibfnamefont{K.~S.} \bibnamefont{Novoselov}},
  \bibinfo{author}{\bibfnamefont{D.}~\bibnamefont{Jiang}},
  \bibinfo{author}{\bibfnamefont{R.}~\bibnamefont{Yang}},
  \bibinfo{author}{\bibfnamefont{T.~J.} \bibnamefont{Booth}}, \bibnamefont{and}
  \bibinfo{author}{\bibfnamefont{A.~K.} \bibnamefont{Geim}},
  \bibinfo{journal}{Applied Physics Letters} \textbf{\bibinfo{volume}{91}},
  \bibinfo{eid}{063124} (\bibinfo{year}{2007}).

\bibitem[{\citenamefont{Ferrari et~al.}(2006)\citenamefont{Ferrari, Meyer,
  Scardaci, Casiraghi, Lazzeri, Mauri, Piscanec, Jiang, Novoselov, Roth
  et~al.}}]{Ferrari2006}
\bibinfo{author}{\bibfnamefont{A.~C.} \bibnamefont{Ferrari}},
  \bibinfo{author}{\bibfnamefont{J.~C.} \bibnamefont{Meyer}},
  \bibinfo{author}{\bibfnamefont{V.}~\bibnamefont{Scardaci}},
  \bibinfo{author}{\bibfnamefont{C.}~\bibnamefont{Casiraghi}},
  \bibinfo{author}{\bibfnamefont{M.}~\bibnamefont{Lazzeri}},
  \bibinfo{author}{\bibfnamefont{F.}~\bibnamefont{Mauri}},
  \bibinfo{author}{\bibfnamefont{S.}~\bibnamefont{Piscanec}},
  \bibinfo{author}{\bibfnamefont{D.}~\bibnamefont{Jiang}},
  \bibinfo{author}{\bibfnamefont{K.~S.} \bibnamefont{Novoselov}},
  \bibinfo{author}{\bibfnamefont{S.}~\bibnamefont{Roth}}, \bibnamefont{et~al.},
  \bibinfo{journal}{Phys. Rev. Lett.} \textbf{\bibinfo{volume}{97}},
  \bibinfo{pages}{187401} (\bibinfo{year}{2006}).

\bibitem[{\citenamefont{Chen et~al.}(2009)\citenamefont{Chen, Jang, Ishigami,
  Xiao, Cullen, Williams, and Fuhrer}}]{Chen2009}
\bibinfo{author}{\bibfnamefont{J.-H.} \bibnamefont{Chen}},
  \bibinfo{author}{\bibfnamefont{C.}~\bibnamefont{Jang}},
  \bibinfo{author}{\bibfnamefont{M.}~\bibnamefont{Ishigami}},
  \bibinfo{author}{\bibfnamefont{S.}~\bibnamefont{Xiao}},
  \bibinfo{author}{\bibfnamefont{W.}~\bibnamefont{Cullen}},
  \bibinfo{author}{\bibfnamefont{E.}~\bibnamefont{Williams}}, \bibnamefont{and}
  \bibinfo{author}{\bibfnamefont{M.}~\bibnamefont{Fuhrer}},
  \bibinfo{journal}{Solid State Communications} \textbf{\bibinfo{volume}{149}},
  \bibinfo{pages}{1080 } (\bibinfo{year}{2009}).

\bibitem[{\citenamefont{Cooper et~al.}(2004)\citenamefont{Cooper, Steffen,
  McDermott, Simmonds, Oh, Hite, Pappas, and Martinis}}]{Cooper2004}
\bibinfo{author}{\bibfnamefont{K.~B.} \bibnamefont{Cooper}},
  \bibinfo{author}{\bibfnamefont{M.}~\bibnamefont{Steffen}},
  \bibinfo{author}{\bibfnamefont{R.}~\bibnamefont{McDermott}},
  \bibinfo{author}{\bibfnamefont{R.~W.} \bibnamefont{Simmonds}},
  \bibinfo{author}{\bibfnamefont{S.}~\bibnamefont{Oh}},
  \bibinfo{author}{\bibfnamefont{D.~A.} \bibnamefont{Hite}},
  \bibinfo{author}{\bibfnamefont{D.~P.} \bibnamefont{Pappas}},
  \bibnamefont{and} \bibinfo{author}{\bibfnamefont{J.~M.}
  \bibnamefont{Martinis}}, \bibinfo{journal}{Phys. Rev. Lett.}
  \textbf{\bibinfo{volume}{93}}, \bibinfo{pages}{180401}
  (\bibinfo{year}{2004}).

\bibitem[{\citenamefont{Martinis et~al.}(2005)\citenamefont{Martinis, Cooper,
  McDermott, Steffen, Ansmann, Osborn, Cicak, Oh, Pappas, Simmonds
  et~al.}}]{Martinis2005}
\bibinfo{author}{\bibfnamefont{J.~M.} \bibnamefont{Martinis}},
  \bibinfo{author}{\bibfnamefont{K.~B.} \bibnamefont{Cooper}},
  \bibinfo{author}{\bibfnamefont{R.}~\bibnamefont{McDermott}},
  \bibinfo{author}{\bibfnamefont{M.}~\bibnamefont{Steffen}},
  \bibinfo{author}{\bibfnamefont{M.}~\bibnamefont{Ansmann}},
  \bibinfo{author}{\bibfnamefont{K.~D.} \bibnamefont{Osborn}},
  \bibinfo{author}{\bibfnamefont{K.}~\bibnamefont{Cicak}},
  \bibinfo{author}{\bibfnamefont{S.}~\bibnamefont{Oh}},
  \bibinfo{author}{\bibfnamefont{D.~P.} \bibnamefont{Pappas}},
  \bibinfo{author}{\bibfnamefont{R.~W.} \bibnamefont{Simmonds}},
  \bibnamefont{et~al.}, \bibinfo{journal}{Phys. Rev. Lett.}
  \textbf{\bibinfo{volume}{95}}, \bibinfo{pages}{210503}
  (\bibinfo{year}{2005}).

\bibitem[{\citenamefont{Palomaki et~al.}(2010)\citenamefont{Palomaki, Dutta,
  Lewis, Przybysz, Paik, Cooper, Kwon, Anderson, Lobb, Wellstood
  et~al.}}]{Palomaki2010}
\bibinfo{author}{\bibfnamefont{T.~A.} \bibnamefont{Palomaki}},
  \bibinfo{author}{\bibfnamefont{S.~K.} \bibnamefont{Dutta}},
  \bibinfo{author}{\bibfnamefont{R.~M.} \bibnamefont{Lewis}},
  \bibinfo{author}{\bibfnamefont{A.~J.} \bibnamefont{Przybysz}},
  \bibinfo{author}{\bibfnamefont{H.}~\bibnamefont{Paik}},
  \bibinfo{author}{\bibfnamefont{B.~K.} \bibnamefont{Cooper}},
  \bibinfo{author}{\bibfnamefont{H.}~\bibnamefont{Kwon}},
  \bibinfo{author}{\bibfnamefont{J.~R.} \bibnamefont{Anderson}},
  \bibinfo{author}{\bibfnamefont{C.~J.} \bibnamefont{Lobb}},
  \bibinfo{author}{\bibfnamefont{F.~C.} \bibnamefont{Wellstood}},
  \bibnamefont{et~al.}, \bibinfo{journal}{Phys. Rev. B}
  \textbf{\bibinfo{volume}{81}}, \bibinfo{pages}{144503}
  (\bibinfo{year}{2010}).

\bibitem[{\citenamefont{Bolotin et~al.}(2008)\citenamefont{Bolotin, Sikes,
  Jiang, Klima, Fudenberg, Hone, Kim, and Stormer}}]{Bolotin2008}
\bibinfo{author}{\bibfnamefont{K.}~\bibnamefont{Bolotin}},
  \bibinfo{author}{\bibfnamefont{K.}~\bibnamefont{Sikes}},
  \bibinfo{author}{\bibfnamefont{Z.}~\bibnamefont{Jiang}},
  \bibinfo{author}{\bibfnamefont{M.}~\bibnamefont{Klima}},
  \bibinfo{author}{\bibfnamefont{G.}~\bibnamefont{Fudenberg}},
  \bibinfo{author}{\bibfnamefont{J.}~\bibnamefont{Hone}},
  \bibinfo{author}{\bibfnamefont{P.}~\bibnamefont{Kim}}, \bibnamefont{and}
  \bibinfo{author}{\bibfnamefont{H.}~\bibnamefont{Stormer}},
  \bibinfo{journal}{Solid State Communications} \textbf{\bibinfo{volume}{146}},
  \bibinfo{pages}{351 } (\bibinfo{year}{2008}).

\bibitem[{\citenamefont{Du et~al.}(2009)\citenamefont{Du, Skachko, Duerr,
  Luican, and Andrei}}]{Du2009}
\bibinfo{author}{\bibfnamefont{X.}~\bibnamefont{Du}},
  \bibinfo{author}{\bibfnamefont{I.}~\bibnamefont{Skachko}},
  \bibinfo{author}{\bibfnamefont{F.}~\bibnamefont{Duerr}},
  \bibinfo{author}{\bibfnamefont{A.}~\bibnamefont{Luican}}, \bibnamefont{and}
  \bibinfo{author}{\bibfnamefont{E.~Y.} \bibnamefont{Andrei}},
  \bibinfo{journal}{Nature} \textbf{\bibinfo{volume}{462}},
  \bibinfo{pages}{192} (\bibinfo{year}{2009}).

\bibitem[{\citenamefont{Bolotin et~al.}(2009)\citenamefont{Bolotin, Ghahari,
  Shulman, Stormer, and Kim}}]{Bolotin2009}
\bibinfo{author}{\bibfnamefont{K.~I.} \bibnamefont{Bolotin}},
  \bibinfo{author}{\bibfnamefont{F.}~\bibnamefont{Ghahari}},
  \bibinfo{author}{\bibfnamefont{M.~D.} \bibnamefont{Shulman}},
  \bibinfo{author}{\bibfnamefont{H.~L.} \bibnamefont{Stormer}},
  \bibnamefont{and} \bibinfo{author}{\bibfnamefont{P.}~\bibnamefont{Kim}},
  \bibinfo{journal}{Nature} \textbf{\bibinfo{volume}{462}},
  \bibinfo{pages}{196} (\bibinfo{year}{2009}).

\bibitem[{\citenamefont{Du et~al.}(2008{\natexlab{b}})\citenamefont{Du,
  Skachko, Barker, and Andrei}}]{Du2008susg}
\bibinfo{author}{\bibfnamefont{X.}~\bibnamefont{Du}},
  \bibinfo{author}{\bibfnamefont{I.}~\bibnamefont{Skachko}},
  \bibinfo{author}{\bibfnamefont{A.}~\bibnamefont{Barker}}, \bibnamefont{and}
  \bibinfo{author}{\bibfnamefont{E.~Y.} \bibnamefont{Andrei}},
  \bibinfo{journal}{Nat Nano} \textbf{\bibinfo{volume}{3}},
  \bibinfo{pages}{491} (\bibinfo{year}{2008}{\natexlab{b}}).

\bibitem[{\citenamefont{Ramos et~al.}(2001)\citenamefont{Ramos, Gubrud,
  Berkley, Anderson, Lobb, and Wellstood}}]{919517}
\bibinfo{author}{\bibfnamefont{R.~C.} \bibnamefont{Ramos}},
  \bibinfo{author}{\bibfnamefont{M.~A.} \bibnamefont{Gubrud}},
  \bibinfo{author}{\bibfnamefont{A.~J.} \bibnamefont{Berkley}},
  \bibinfo{author}{\bibfnamefont{J.~R.} \bibnamefont{Anderson}},
  \bibinfo{author}{\bibfnamefont{C.~J.} \bibnamefont{Lobb}}, \bibnamefont{and}
  \bibinfo{author}{\bibfnamefont{F.~C.} \bibnamefont{Wellstood}},
  \bibinfo{journal}{IEEE Trans. Appl. Supercond.}
  \textbf{\bibinfo{volume}{11}}, \bibinfo{pages}{998} (\bibinfo{year}{2001}).

\bibitem[{\citenamefont{Xu}(2004)}]{Xuthesis}
\bibinfo{author}{\bibfnamefont{H.}~\bibnamefont{Xu}}, Ph.D. thesis,
  \bibinfo{school}{University of Maryland College Park} (\bibinfo{year}{2004}).

\bibitem[{\citenamefont{Titov and Beenakker}(2006)}]{Titov2006}
\bibinfo{author}{\bibfnamefont{M.}~\bibnamefont{Titov}} \bibnamefont{and}
  \bibinfo{author}{\bibfnamefont{C.~W.~J.} \bibnamefont{Beenakker}},
  \bibinfo{journal}{Phys. Rev. B} \textbf{\bibinfo{volume}{74}},
  \bibinfo{pages}{041401} (\bibinfo{year}{2006}).

\bibitem[{\citenamefont{Washburn et~al.}(1985)\citenamefont{Washburn, Webb,
  Voss, and Faris}}]{washburn1985}
\bibinfo{author}{\bibfnamefont{S.}~\bibnamefont{Washburn}},
  \bibinfo{author}{\bibfnamefont{R.~A.} \bibnamefont{Webb}},
  \bibinfo{author}{\bibfnamefont{R.~F.} \bibnamefont{Voss}}, \bibnamefont{and}
  \bibinfo{author}{\bibfnamefont{S.~M.} \bibnamefont{Faris}},
  \bibinfo{journal}{Phys. Rev. Lett.} \textbf{\bibinfo{volume}{54}},
  \bibinfo{pages}{2712} (\bibinfo{year}{1985}).

\bibitem[{\citenamefont{Tinkham}(1975)}]{Tinkham}
\bibinfo{author}{\bibfnamefont{M.}~\bibnamefont{Tinkham}},
  \emph{\bibinfo{title}{Introduction to Superconductivity}}
  (\bibinfo{publisher}{McGraw-Hill}, \bibinfo{address}{New York},
  \bibinfo{year}{1975}).

\bibitem[{\citenamefont{Wallraff et~al.}(2003)\citenamefont{Wallraff, Duty,
  Lukashenko, and Ustinov}}]{Wallraff2003}
\bibinfo{author}{\bibfnamefont{A.}~\bibnamefont{Wallraff}},
  \bibinfo{author}{\bibfnamefont{T.}~\bibnamefont{Duty}},
  \bibinfo{author}{\bibfnamefont{A.}~\bibnamefont{Lukashenko}},
  \bibnamefont{and} \bibinfo{author}{\bibfnamefont{A.~V.}
  \bibnamefont{Ustinov}}, \bibinfo{journal}{Phys. Rev. Lett.}
  \textbf{\bibinfo{volume}{90}}, \bibinfo{pages}{037003}
  (\bibinfo{year}{2003}).

\end{thebibliography}

\end{document}